\documentclass[a4paper]{jpconf}
\usepackage{epsfig, cite}
\usepackage{graphicx}

\newcommand{\charge}{q}

\newcommand{\be}[1]{\begin{equation} #1 \end{equation}}

\newcommand{\paren}[1]{\left( #1 \right)}

\newcommand{\dis}{\displaystyle}
\newcommand{\goodspace}{\hspace{0.6cm}}

\def\apriori{{\it a priori\/}}

\def\eg{{\it e.g.\ }}

\def\ie{{\it i.e.\ }}

\begin{document}

{\hfill MIT-CTP 3787} 
\newline
\vspace{0.1cm}
{\hfill QMUL-PH-06-09}
\newline
\vspace{0.1cm}
{\hfill USITP 06-04}

\title{Ruppeiner theory of black hole thermodynamics\footnote{Talk given by Narit Pidokrajt at the XXIX Spanish Relativity Meeting (ERE2006), Palma de Mallorca, Spain}}

\author{Jan E \AA man$^1$, James Bedford$^{2, 3}$, Daniel~Grumiller$^4$, Narit Pidokrajt$^1$ and John Ward$^{2,3}$}

\address{$^1$ Fysikum, AlbaNova, Stockholm University, SE-106 91 Stockholm, Sweden}

\address{$^2$ Department of Physics, Queen Mary, University of London, Mile End Road, London E1~4NS, United Kingdom}

\address{$^3$ Theory Division, CERN, CH-1211 Geneve 23, Switzerland}

\address{$^4$ Center for Theoretical Physics, Massachusetts Institute of Technology, 77~Massachusetts~Ave., Cambridge, MA 02139, USA }

%\ead{narit@physto.se}

\ead{ja@physto.se, j.a.p.bedford@qmul.ac.uk, grumil@hep.itp.tuwien.ac.at,  narit@physto.se, j.ward@qmul.ac.uk}

\begin{abstract}

The Ruppeiner metric as determined by the Hessian of the Gibbs surface provides a geometric description of thermodynamic systems in equilibrium. An interesting example is a black hole in equilibrium with its own Hawking radiation. In this article, we present results from the Ruppeiner study of various black hole families from different gravity theories \eg 2D dilaton gravity, BTZ, general relativity and higher-dimensional Einstein-Maxwell gravity.

\end{abstract}

\section{Introduction --- Ruppeiner: thermodynamics as geometry}

In 1979 George Ruppeiner proposed a geometrical way of studying thermodynamics of equilibrium systems \cite{ruppeiner1}.
In his theory certain aspects of thermodynamics and statistical mechanics of the system under consideration
are encoded in a single geometrical object, \ie the metric describing its thermodynamic state space. The distance on this
space is given by
\be{
ds^2_R = g^R_{ij} dX^i dX^j,
}
where $g^R_{ij}$ is the so-called Ruppeiner metric defined as
\be{
g^R_{ij} = - \partial_i \partial_j S(X), \,\,\,  X = (U, N^a); \,\, a= 1,2, \ldots, n
}
where $U$ is the system's internal energy (which is the mass in black hole thermodynamics), $S(X)$ is an entropy function of the thermodynamic system one wishes to consider and $N^a$ stand for other extensive variables, or---in our application---mechanically conserved charges of the system. It was observed by Ruppeiner that in thermodynamic fluctuation theory the Riemannian curvature\footnote{Henceforth we refer to it as the Ruppeiner curvature.} of the Ruppeiner metric measures the complexity of the underlying statistical mechanical model, \ie it is flat for the ideal gas whereas any curvature singularities are a signal of critical behavior.  The Ruppeiner theory has been applied to numerous systems and yielded significant results, for details see \cite{ruppeiner2}. Mathematically the Ruppeiner geometry is one particular type of information geometry. Ruppeiner originally developed his theory in the context of thermodynamic fluctuation theory, for systems in canonical ensembles. Most black holes have negative specific heats and are described microcanonically. In spite of this we have found that the Ruppeiner geometry of black holes is often surprisingly simple and elegant. Furthermore some findings are physically suggestive.

The Ruppeiner geometry is conformally related to the so-called Weinhold geometry \cite{weinhold} as
\be{
\label{eq:conformal}
g^W_{ij} = T g^R_{ij},
}
where $T$ is the temperature of the system under consideration. However the Weinhold metric is originally defined as
the Hessian of the energy (mass) with respect to entropy and other extensive parameters, namely
\be{
g^W_{ij} = \partial_i \partial_j U (S, N^a).
}
The Weinhold geometry does not have the same physical meaning as the Ruppeiner geometry but we have found it very useful in most of our calculations when the Ruppeiner geometry is not easily tractable.

Most of our presentation has the character of a review, with details of the calculations appearing in the cited literature. The results in section \ref{se:dil} are new and thus presented in somewhat more detail.

\section{Applications to black hole thermodynamics}

The Ruppeiner geometries\footnote{Also known as thermodynamic geometries.}
of black hole  (BH) thermodynamics have been worked out for a number of
BH families by several groups of people \cite{china1, ourpaper1,
israel, ourpaper2, china2, ourpaper3, india,  ourpaper4, ourpaper5,
ourpaper6}. It is certainly natural to start off with the BH
solutions in general relativity (4D) as done in \cite{ourpaper1}. We have
found that the geometrical pattern is very similar in the higher
dimensional spacetimes as well \cite{ourpaper2}. It is also sensible to
apply this theory to the BH solutions in lower dimensions such as
2D dilaton gravity \cite{ourpaper6} as well as the BTZ BH
\cite{ourpaper1, india}. In Ref.~\cite{israel} the Ruppeiner geometry was used to address the issues
of stability, divergence of fluctuations and critical phenomena in the BH/black ring system. 
  Furthermore, it is attractive to apply this theory to low-energy string BHs (such as the dilaton BH in 4D with a unit coupling constant)~\cite{ourpaper4}. We have also been attempting to work out the Ruppeiner geometry of the BH
solutions in the AdS background where the cosmological constant is treated
as a parameter of the theory \cite{ourpaper5}. In this section we briefly
show the calculations done for some BH solutions, \ie  2D
dilaton gravity, the BTZ BH, Reissner-Nordstr\"om (RN) and Kerr BHs in arbitrary dimension \cite{ourpaper2} and the 4D Einstein-Maxwell-dilaton BH \cite{ourpaper4}. Finally, we have constructed a mathematical flatness theorem \cite{ourpaper3} which gives a condition for the Ruppeiner geometry to be flat. It applies to many BH families studied so far.

\subsection{2D dilaton gravity}\label{se:dil}

We consider here 2D dilaton gravity (for a comprehensive review cf.~e.g.~\cite{Grumiller:2002nm}),
\begin{equation}
 \label{eq:rup1}
 I_{2DG}=\frac{1}{4\pi}\int \: d^2  x\sqrt{-g}\left(XR+U(X)(\nabla X)^2-2V(X,\charge)\right)\,,
\end{equation}
where $X$ is a scalar field (dilaton), $U,V$ are arbitrary functions thereof defining the model and $R$ is the Ricci scalar associated with the 2D metric $g_{\mu\nu}$. The function $V$ additionally depends on a parameter $\charge$ which may be interpreted as charge.\footnote{Such a dependence on $\charge$ emerges for instance if one introduces in 2D an abelian Maxwell-term and integrates it out exactly. Its only remnant is the conserved $U(1)$ charge $\charge$ which enters the potential $V$.} In this way charged BH solutions can be described such as the RN BH. We shall employ the definitions
\begin{equation}
  \label{eq:rup5}
  Q(X)=\int^XU(z) d z\,,\qquad w(X,\charge)=\int^X e^{Q(z)}V(z,\charge)d z\,.
\end{equation}
The quantity $w(X,\charge)$ is invariant under dilaton dependent conformal transformations. In terms of these functions it can be shown that the solution for the line-element in Eddington-Finkelstein gauge reads
\begin{equation}
  \label{eq:rup6}
  d s^2=2e^{Q(X)} d u\left(d X - (w(X,\charge)+M)d u\right)\,,
\end{equation}
where $M$ is a constant of motion corresponding to the mass. Killing horizons emerge for
\begin{equation}
  \label{eq:rup7}
  w(X,\charge)+M=0\,.
\end{equation}
The solution of this equation for the outermost horizon is denoted by $X=X_h$.
The Hawking-Unruh temperature (as derived e.g.~from surface gravity \cite{Kummer:1999zy}) is given by
\begin{equation}
  \label{eq:rup2}
  T=|w'(X,\charge)|_{X=X_h}\,.
\end{equation}
Prime denotes differentiation with respect to $X$. The Bekenstein-Hawking entropy (as derived e.g.~from Wald's Noether charge technique \cite{Gegenberg:1994pv}) is given by
\begin{equation}
  \label{eq:rup3}
  S=X_h\,.
\end{equation}
We are now able to derive the Weinhold and Ruppeiner metrics. Because of (\ref{eq:rup7})-(\ref{eq:rup3}) both metrics depend on the conformally invariant function $w(X,\charge)$ only. So we are free to choose $Q=0$ to simplify the calculations. Putting together all definitions yields the Weinhold metric
\begin{equation}
  \label{eq:weinhold2DG}
  d s^2_W=-w''(S,\charge) d S^2-2\dot{w}^\prime(S,\charge) d S d\charge-\ddot{w}(S,\charge)d\charge^2\,,
\end{equation}
where dot denotes differentiation with respect to $\charge$. The Ruppeiner metric follows as
\begin{equation}
  \label{eq:ruppeiner2DG}
   d s^2_R=\frac{1}{|w'(S,\charge)|}d s^2_W\,.
\end{equation}
The conformal factor between these two metrics never vanishes unless the horizon degenerates. We discuss briefly two important classes of examples and refer to \cite{ourpaper6} for an extensive discussion.

\paragraph{Reissner-Nordstr\"om like BHs} The family of models ($b\neq -1\neq c$)
\begin{equation}
  \label{eq:rup4}
  w=-\frac{A}{b+1}X^{b+1}-\frac{B}{2(c+1)}X^{c+1}\charge^2
\end{equation}
is simple and interesting, as it contains the spherically reduced RN BH from $D$ dimensions $b=-1/(D-2)$, as well as charged versions of the Witten BH $b=0$ \cite{WittenBH} and of the Jackiw-Teitelboim model $b=1$ \cite{JT}. With the coordinate redefinition $u=\charge S^{c+1}$ the Weinhold metric simplifies to diagonal form,
\begin{equation}
  \label{eq:rup8}
  d s^2_W=(bAS^{b-1}-(\frac c2+1)Bu^2S^{-c-3})d S^2+\frac{B}{c+1}S^{-c-1}d u^2\,.
\end{equation}
It is flat for $b=0$ or $c=b-2$. Similarly, the Ruppeiner metric turns out as 
\begin{equation}
  \label{eq:rup9}
  d s^2_R=\frac{1}{S(AS^b+\frac B2 u^2 S^{-c-2})}\left[b(AS^b-\frac{\frac c2+1}{b}Bu^2S^{-c-2})d S^2+\frac{B}{c+1}S^{-c}d u^2\right]\,.
\end{equation}
The Ruppeiner metric (\ref{eq:rup9}) is not flat in general. However, if the condition $c=-b-2$ holds, then (\ref{eq:rup9}) simplifies considerably,
\begin{equation}
  \label{eq:rup10}
   d s^2_R=b \frac{d S^2}{S}+2S\frac{1}{(b+1)}\frac{d u^2}{(-2\frac AB - u^2)}\,.
\end{equation}
The Ruppeiner metric (\ref{eq:rup10}) is flat and has Lorentzian or Euclidean signature, depending on $b$ and the sign of $u^2+2A/B$. The particular subclass
\begin{equation}
  \label{eq:rup11}
  U=-\frac{b+1}{X}\,,\qquad V=-AX^{2b+1} -\frac{B}{2}\frac{q^2}{X}
\end{equation}
describes the spherically reduced RN BH from $D$ dimensions with $b=-1/(D-2)$. It fulfills the condition $c=-b-2$, and thus all corresponding Ruppeiner metrics are flat. This agrees with the results in section \ref{se:RN} below: the line-element (\ref{eq:rup10}) essentially coincides with the line-element (\ref{eq:RuppeinerRN-D}) upon rescaling $u$ and choosing $B$ appropriately.

\paragraph{Chern-Simons like BHs} In some cases, like the Kaluza-Klein reduced gravitational Chern-Simons term \cite{Guralnik:2003we} or the toroidally reduced
BTZ BH \cite{Achucarro:1993fd}, the charge $q$ does not enter quadratically in the potential but only linearly. Therefore, we consider here the class of models defined by
\begin{equation}
  \label{eq:rup12}
  w=-\frac{A}{b+1}X^{b+1}-\frac{B}{c+1}X^{c+1}\charge\,.
\end{equation}
We obtain the flat\footnote{The line element (\ref{eq:rup13}) describes a flat (Rindler type) geometry because the coordinate $q$ appears only linearly.} Weinhold metric
\begin{equation}
  \label{eq:rup13}
  d s^2_W = (bAS^{b-1}+cBS^{c-1}\charge)d S^2+2BS^cd Sd\charge\,,
\end{equation}
and the Ruppeiner metric 
\begin{equation}
  \label{eq:rup14}
  d s^2_R = \frac{1}{AS^b+B\charge S^c}\left[\frac{b}{S}(AS^b+\frac cb B\charge S^c)d S^2+2BS^cd Sd\charge\right]
\end{equation}
The Ruppeiner metric (\ref{eq:rup14}) is not flat in general. However, if the condition $c=b$ holds, then (\ref{eq:rup14}) simplifies considerably,
\begin{equation}
  \label{eq:rup15}
   d s^2_R=b \frac{d S^2}{S}+B\frac{2d Sd\charge}{A+B\charge}\,.
\end{equation}
The Ruppeiner metric (\ref{eq:rup15}) is flat and has Lorentzian signature.
%\paragraph{} 

We conclude this section with a remark on a duality found recently \cite{Grumiller:2006xz}. It connects two different models leading to the same classical solutions for the line-element (\ref{eq:rup6}) and therefore to the same surface gravity (\ref{eq:rup2}), but the respective entropies differ in general. It would be interesting to study the behavior of the Weinhold and Ruppeiner metrics under this duality.

\subsection{BTZ black hole}

The BTZ BH \cite{btz} occurs as a solution of 2+1 gravity and a toy model of gravity in 3+1 dimensions. It is very important for studies in BH physics \eg it is a counterpart of the Kerr BH in 4D.  We have found that the BTZ BH has thermodynamic geometries similar to that of the RN BH, \ie it has a flat Ruppeiner geometry and curved Weinhold geometry except that they have opposite metric signatures. The pleasant fact about this BH is that all its heat capacities are positive, yet it is a \emph{bona fide} BH. We investigate the thermodynamic geometry \cite{ourpaper1} of this BH by starting with the Weinhold metric as it is simpler, \ie we start off the with mass function given as
\be{
M = S^2 + \frac{J^2}{4S^2}.
}
The Weinhold metric for the BTZ BH, 
\be{
ds^2_W = \paren{2 + \frac{3J^2}{2S^4}} dS^2 - \paren{\frac{2J}{S^3}} dS dJ + \paren{\frac{1}{2S^2}} dJ^2,
}
is non-flat. Using the conformal relation (\ref{eq:conformal}) we obtain the Ruppeiner metric, which after diagonalization, is given by
\be{
ds^2_R = \frac{1}{S}dS^2 + \paren{\frac{S}{1 - u^2}}du^2,
\label{eq:BTZ-Ruppeiner}
}
where we have used
\be{
u = \frac{J}{2S^2}; \goodspace -1 \leq u \leq 1.
}
The Ruppeiner metric in (\ref{eq:BTZ-Ruppeiner}) is flat, in other words the space of its thermodynamic states is flat. This can be shown to be a wedge of a Euclidean flat space. The Ruppeiner geometry of the BTZ BH has also been studied elsewhere \cite{china1, india}. Note also that the Euclidean signature of the Ruppeiner metric corresponds to all heat capacities being positive.

In Ref. \cite{india} the BTZ BH in the presence of the Chern-Simons (CS) term \cite{chern-simons} was investigated. The authors found that Ruppeiner geometry is still flat. The result without the CS term agrees with ours. They have also studied the entropy of the BTZ BH with the logarithmic correction which gives rise to a nonzero curvature scalar. See Table \ref{table:result1} for a summary on the Ruppeiner theory of BHs in lower-dimensional spacetime.

\subsection{Reissner-Nordstr\"om black hole in arbitrary dimension}\label{se:RN}

In this section we present the Ruppeiner study of the RN BH in arbitrary spacetime \cite{myers-perry} as it is a direct generalization  of that in 4D. The entropy in terms of the BH's mass and electric charge is given by
\be{
S = \paren{M + M\sqrt{1 - \frac{D-2}{2(D-3)} \frac{Q^2}{M^2}}}^{\frac{D-2}{D-3}}.
}
We have learned from our previous work that for the RN BH it is simpler to begin with the Weinhold metric, hence
we invert the entropy equation and obtain
\be{
M = \frac{S^{\frac{D-3}{D-2}}}{2} + \frac{D-2}{2(D-3)} \frac{Q^2}{S^{\frac{D-3}{D-2}}}.
}
The Weinhold metric of the RN BH in arbitrary dimension can be diagonalized by choosing the new coordinate
\be{
u = \sqrt{\frac{D-2}{2(D-3)}} \frac{Q}{ S^{\frac{D-3}{D-2}} }; \goodspace -1 \leq u \leq 1.
}
The Weinhold metric now becomes
\be{
ds^2_W =  S^{\frac{D-1}{D-2}}\paren{ -\frac{1}{2} \frac{D-3}{(D-2)^2} (1-u^2)dS^2 + S^2 du^2  },
}
which is a curved Lorentzian metric. The Ruppeiner metric can be obtained by using the conformal relation, thus
\be{
ds^2_R = \frac{-1}{D-2}\frac{dS^2}{S} + 2S \frac{D-3}{D-2} \frac{du^2}{1 - u^2}.
\label{eq:RuppeinerRN-D}
}
This is a flat metric. It can be written in Rindler coordinates by using
\begin{figure}
\centering
\psfig{file=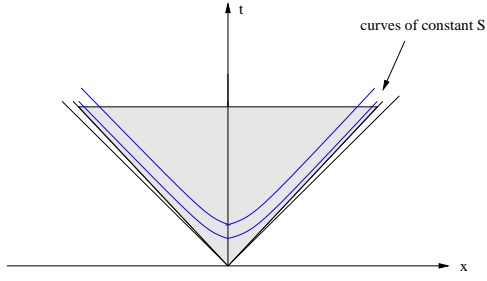,width=.4\textwidth}
\caption{The state space of the 4D RN BHs shown as a wedge in a flat Minkowski space. }
\label{fig:plotRN}
\end{figure}
\be{
\tau = 2\sqrt{\frac{S}{D-2}} \goodspace \textrm{and} \goodspace u = \sin \frac{\sigma\sqrt{2(D-3)}}{D-2}.
}
The angle $\sigma$ then lies within the following interval
\be{
-\frac{(D-2)\pi}{2\sqrt{2(D-3)}}  \leq  \sigma \leq  \frac{(D-2)\pi}{2\sqrt{2(D-3)}}.
}
Turning this into Minkowski coordinates $(t,x)$ we obtain
\be{
ds^2 = -d\tau^2 + \tau^2 d\sigma^2 = -dt^2 + dx^2,
}
where we have used
\be{
t = \tau \cosh \sigma \goodspace \textrm{and} \goodspace x = \tau \sinh \sigma.
}
The Lorentzian signature of the Ruppeiner metric corresponds to some heat capacities being negative. Using the new parameters defined in terms of mass and charge, we can represent the entropy of the RN BH in Minkowskian coordinates as follows:
\be{
S = \frac{1}{4}(D-2)(t^2 - x^2).
}
The Ruppeiner metric can be presented as a Rindler wedge as shown in Fig \ref{fig:plotRN}. Note that curves of constant $S$ are segments of hyperbolas and the opening angle of the wedge grows as $D \rightarrow \infty$. For the RN BH in the AdS background, its Ruppeiner and Weinhold geometries are non-flat. However some physical significance can be extracted from the Ruppeiner curvature in that the signature of the metric together with its stability properties changes for sufficiently large black holes, which is a well-known feature \cite{hawking-page}.  The Ruppeiner curvature of the RN AdS BH is singular in the extremal limit and along the curve where the Ruppeiner metric changes signature. 

\subsection{Kerr black hole in arbitrary dimension}

In $D$ dimensions, the mass of the Kerr BH is given by 
\be{
M(S, J ; D) = \frac{D-2}{4}S^{\frac{D-3}{D-2}} \paren{1 + \frac{4J^2}{S^2}}^{1/(D-2)},
}
therefore it is possible to compute the Weinhold geometry directly. With the conformal relation, we can obtain the Ruppeiner metric which is found to be curved. The Ruppeiner scalar of the Kerr BH in arbitrary dimension is given by
\be{
\dis
R =  - \frac{1}{S} \frac{\dis 1 - 12\frac{D-5}{D-3}\frac{J^2}{S^2}}
{\dis \paren{1 - 4\frac{D-5}{D-3} \frac{J^2}{S^2}}\paren{1 + 4\frac{D-5}{D-3} \frac{J^2}{S^2}}}.
\label{eq:DKerr-curvature}
}
In 4D the curvature scalar diverges along the curve $4J^2 = S^2$ consistent with the previous result \cite{ourpaper1}.  Remarkably, in $D > 5$ we have a curvature divergence but not in the limit of extremality, rather at
\be{
4J^2 = \frac{D-3}{D-5} S^2.
}
This is where Emparan and Myers \cite{emparan} suggest that the Kerr BH becomes unstable and changes its behavior to be like a black membrane. This is also where the temperature of the higher-dimensional ($D \geq 6$) Kerr BH reaches its minimum. The wedge of the state space of the Kerr BH in $D >  5$ fills the entire null cone, owing to the fact that for such a BH there is no genuine extremal limit.  Additionally, in $D \geq 5$ the Kerr BH can have more than one angular momentum. We did study the Ruppeiner geometry of this BH in 5D where the BH can have two angular momenta. Both Weinhold and Ruppeiner metrics of the double-spin 5D Kerr BH are non-flat.

Furthermore, in 5D for some values of the parameters, there exist black ring solutions \cite{blackring} whose entropy is larger than that of the BH studied in this paper. However according to (\ref{eq:DKerr-curvature}) nothing special happens to the Gibbs surface of the Kerr BH \emph{\`{a} la} Myers-Perry. Full details on the Ruppeiner study of Kerr and Kerr double-spin BHs can be found in Ref.~\cite{ourpaper2}.  See Table \ref{table:result3} for a summary of the Ruppeiner theory of BHs in higher-dimensional spacetime.

\subsection{Dilaton black hole}

The entropy of the dilaton BH in 4D with arbitrary constant $a$ is given by
\be{
S =  M^2\paren{ 1 + \sqrt{1 - (1-a^2)\frac{Q^2}{M^2}}}^2\paren{1 -  \frac{\dis (1 + a^2)Q^2}{M^2\paren{\dis 1 + \sqrt{1 - (1-a^2)\frac{Q^2}{M^2}}}^2}}^{\frac{2a^2}{1 + a^2}}.
\label{eq:dilaton-entropy}
}
In the case of $a=1$ and a few others we are able to compute both the Ruppeiner and Weinhold geometries exactly.  As anticipated, the state space of this BH changes with $a$ in that the state space of the dilaton BH with $a>0$ lies on the null cone (where the entropy vanishes). For arbitrary $a$ the Ruppeiner geometry is flat as shown by the flatness theorem discussed below. The Weinhold geometry of this BH is found to be curved.

\subsection{The flatness theorem}

In Ref.~\cite{ourpaper3} we have found a sufficient but not necessary condition for the Ruppeiner (Weinhold) geometry to be flat for a certain class of BH solutions. The underlying mechanism for the flatness has to do with the scale-invariance of the Einstein-Maxwell action. More information and a mathematical proof can be found in the reference but in short we can state the theorem as follows:

If the entropy of the BH is given by
\be{
\label{eq:flatness}
S = M^a f(M^b Q)
}
where $M$ is the BH's mass and $Q$ the conserved charge, the Ruppeiner metric will be a flat metric if $b=-1$ and $a \neq1$. This theorem can be applied directly to the cases of BTZ, BTZ-CS, RN and dilaton BH in 2D and 4D.

It may be worth noting that the Ruppeiner geometry of the ideal gas is flat for a different mathematical reason\footnote{That is, the ideal gas has a flat Ruppeiner geometry despite the fact that its entropy does not obey (\ref{eq:flatness}). In general the entropy function of the ideal gas is a homogeneous function of three parameters, $S(U, V, N)$ and the Ruppeiner metric is a degenerate metric unless we keep one of the parameters constant.}. In~\cite{idealgas} it was shown that the ideal gas at constant particle number and constant volume has flat Ruppeiner and Weinhold geometries. The Ruppeiner geometry in both cases have a state space which describes a flat plane. 

\paragraph{} We would like to end this section by mentioning the ongoing work in Ref.~\cite{ourpaper5} which is a study of the thermodynamic geometries of the BHs in the AdS background where the cosmological constant is treated as a parameter of the theory. The flatness (non-flatness) of the Ruppeiner (and Weinhold) geometry does appear to be sensitive to the (re)definition\footnote{In Ref.~\cite{ourpaper5} we use $\Lambda$ (a cosmological constant) or $l = \sqrt{-3/\Lambda}$ as a thermodynamic variable of the theory in both the Ruppeiner and Weinhold metrics and obtain qualitatively distinct results.} of the cosmological constant which is \apriori~unexpected.

\section{More results}

We summarize results obtained from applications of the Ruppeiner theory to BH thermodynamics by us \cite{ourpaper1, ourpaper2, ourpaper3, ourpaper4, ourpaper5, ourpaper6} and others \cite{china1, israel, china2, india} in Tables 1-3. We would like to emphasize that especially for state spaces with dimension three or higher it is convenient to employ algebra programs such as CLASSI \cite{classi} and GRTensor \cite{grtensor}.

\begin{table}[h]
\caption{\label{table:result1} Thermodynamic geometry in lower-dimensional spacetime}
\begin{center}
\begin{tabular}{lll}
\br
BH family & Ruppeiner geometry & Weinhold geometry   \\
\mr
(1+1) RN like BH (generic) &     curved   &       curved \\
(1+1) reduced RN BH ($c=-b-2$) &   flat  &   curved \\
(1+1) CS like BH (generic)  &   curved   &       flat \\
(2+1) BTZ          & flat      & curved      \\
(2+1) BTZ (Chern-Simons) & flat   &  curved  \\
(2+1) BTZ (Log corrections) & curved &  curved  \\
\br
\end{tabular}
\end{center}
\end{table}

\begin{table}[h]
\caption{\label{table:result2} Thermodynamic geometry in four-dimensional spacetime}
\begin{center}
\begin{tabular}{lll}
\br
BH family & Ruppeiner geometry & Weinhold geometry   \\
\mr
 Reissner-Nordstr\"om   &  flat  & curved  \\
 Reissner-Nordstr\"om AdS   & curved   & curved    \\
 Kerr & curved & flat  \\
 Kerr AdS  & curved & curved  \\
 Kerr-Newman & curved & curved \\
\br
\end{tabular}
\end{center}
\end{table}

\begin{table}[h]
\caption{\label{table:result3}  Thermodynamic geometry in higher-dimensional spacetime }
\begin{center}
\begin{tabular}{lll}
\br
BH family & Ruppeiner geometry & Weinhold geometry   \\
\mr
 Reissner-Nordstr\"om   &  flat  & curved \\
 Kerr & curved & flat   \\
 Double-spin Kerr (5D) &  curved  & curved   \\
 Black ring (5D) & curved & flat \\
\br
\end{tabular}
\end{center}
\end{table}

\section{Summary and Outlook}

The Ruppeiner theory is a geometrical theory of thermodynamics. It provides an alternative and elegant
route to obtain insight into thermodynamics through Riemannian geometry. Its power is due to the fact that the Ruppeiner metric together with its associated curvature and signature encodes many aspects of thermodynamics consistent with the known results in the literature.

Since the underlying statistical mechanics of BHs is still
unsettled, the application of the Ruppeiner theory to BH
thermodynamics then gives a new perspective  on this
subject.  Although currently we have few results that are physically suggestive, the geometrical patterns 
we have observed (See Tables 1-3) may play an important role in the future, when quantum gravity is better understood.
As a matter of fact, the Ruppeiner theory of BH thermodynamics can be applied
to every class of BHs as long as their thermodynamic quantities are well-defined. 
The difficulty consists in finding reasonable interpretation(s) of
the calculated Ruppeiner geometries.

%Finally, we must admit that we have been unable to find any simple or telling results for the systems with three dimensional state spaces, despite the fact that many interesting BH families do have three dimensional state spaces.

\section*{Acknowledgments}

NP is supported by Doktorandtj\"anst of Stockholm University and would like to acknowledge the local
organizers (Universitat de les Illes Balears) of the ERE2006 for all their efforts and support. DG is supported by the Marie Curie Fellowship MC-OIF 021421 of the European Commission under the Sixth EU Framework Programme for Research and Technological Development (FP6). This work is supported in part by funds provided by the U.S. Department of Energy (DOE) under the cooperative research agreement DEFG02-05ER41360.

Many thanks go to Ingemar Bengtsson for his critical reading and kind assistance on the fine-tuning of the manuscript. NP appreciates the cordialness of the following participants of the ERE2006 \ie Christian Teijon, Charline Filloux, Antonio Francisco Garc\'ia-Mar\'in, Alba Gutierrez and Norbert Van den Bergh. In particular he is thankful to Isabella Malmn\"as and Tobias Fischer and for making the trip to Mallorca a very pleasant and memorable one.


\section*{References}


\begin{thebibliography}{99}

\bibitem{ruppeiner1} Ruppeiner G 1979 {\it Phys. Rev.} A \textbf{20} 1608

\bibitem{ruppeiner2} Ruppeiner G 1995 {\it Rev. Mod. Phys.} \textbf{67} 605 

\bibitem{weinhold}  Weinhold F 1975 {\it J. Chem. Phys.} \textbf{63} 2479 

\bibitem{china1}  Cai R G and Cho J H 1999 {\it Phys. Rev. D} \textbf{60}  067502 

\bibitem{ourpaper1} \AA man J E, Bengtsson I and Pidokrajt N 2003 {\it Gen. Rel. Grav.} \textbf{35} 1733 

\bibitem{israel} Arcioni G and Lozano-Tellechea E 2005 {\it Phys. Rev. D} {\bf 72} 104021

\bibitem{ourpaper2} \AA man J E and Pidokrajt N 2006 {\it Phys. Rev. D} \textbf{73} 024017 

\bibitem{china2} Shen J, Cai R G, Wang B and Su R K 2005 Thermodynamic Geometry and Critical Behavior of Black Holes {\it Preprint} gr-qc/0512035

\bibitem{ourpaper3} \AA man J E, Bengtsson I and Pidokrajt N 2006 {\it Gen. Rel. Grav.} \textbf{38} 1305

\bibitem{india} Sarkar T, Sengupta G and Tiwari B N 2006 On the Thermodynamic Geometry of BTZ Black Holes {\it Preprint}  hep-th/0606084

\bibitem{ourpaper4} \AA man J E, Bengtsson I, Pidokrajt N and Ward J, in preparation.

\bibitem{ourpaper5} \AA man J E, Bedford J,  Pidokrajt N and Ward J, in preparation.

\bibitem{ourpaper6} \AA man J E, Grumiller D and Pidokrajt N, in preparation.

\bibitem{Grumiller:2002nm}
  Grumiller D, Kummer W and Vassilevich D V 2002 {\it Phys. Rept.} {\bf 369} 327 ({\it Preprint} hep-th/0204253)
\item[] Grumiller D and Meyer R 2006 Ramifications of lineland  ({\it Preprint} hep-th/0604049)

%\cite{Kummer:1999zy}
\bibitem{Kummer:1999zy}
  Kummer W and Vassilevich D V 1999 {\it  Annalen Phys.} {\bf 8} 801 ({\it Preprint} gr-qc/9907041)

\bibitem{Gegenberg:1994pv}
  Gegenberg J, Kunstatter G and Louis-Martinez D 1995 {\it Phys. Rev. D} {\bf 51} 1781 ({\it Preprint} gr-qc/9408015)

\bibitem{WittenBH}
Witten E 1991 {\it Phys. Rev. D} {\bf 44} 314 
\item[]Callan Jr C G, Giddings S B, Harvey J A and Strominger A 1992 {\it Phys. Rev. D} {\bf 45} 1005 ({\it Preprint} hep-th/9111056)

\bibitem{JT}
Jackiw R and Teitelboim C 1984 {\it Quantum Theory Of Gravity}, ed Christensen S (Bristol: Adam Hilger)

\bibitem{Guralnik:2003we}
Guralnik G, Iorio A, Jackiw R and Pi S Y 2003 {\it Annals Phys. } {\bf 308} 222 ({\it Preprint}  hep-th/0305117)
\item[] Grumiller D and Kummer W 2003 {\it Annals Phys.} {\bf 308} 211 ({\it Preprint} hep-th/0306036)
\item[] Bergamin L, Grumiller D, Iorio A and Nu{\~n}ez C 2004  JHEP {\bf 0411} 021  ({\it Preprint} hep-th/0409273)

\bibitem{Achucarro:1993fd}
 Ach\'{u}carro A and Ortiz M E 1993 {\it Phys. Rev. D} {\bf 48} 3600  ({\it Preprint}  hep-th/9304068)

\bibitem{Grumiller:2006xz}
  Grumiller D and Jackiw R 2006 {\it Phys. Lett. B} {\bf 642} 530 ({\it Preprint}  hep-th/0609197)

\bibitem{btz}  Ba\~{n}ados M, Teitelboim C and Zanelli J 1992 {\it Phys. Rev. Lett. } \textbf{69}  1849 

\bibitem{chern-simons} Deser S, Jackiw R and 't Hooft G  1984 {\it Annals Phys.} {\bf 152} 220 

\bibitem{myers-perry} Myers R C and Perry M J 1986 {\it Annals Phys.} \textbf{172} 304 

\bibitem{dilaton} Gibbons G W 1982 {\it Nucl. Phys. B} \textbf{207} 337
\item[] Gibbons G W and Maeda K 1988 {\it Nucl. Phys. B} \textbf{298} 741
\item[] Garfinkle D, Horowitz G T and Strominger A 1991 {\it  Phys. Rev. D} \textbf{43} 3140 
\item[] Horne J H and Horowitz G T 1992 {\it Phys. Rev. D} \textbf{46} 1340 
\item[] Horne J H and Horowitz G T 1993 {\it Nucl. Phys. B} \textbf{399}  169 

\bibitem{emparan} Emparan R and Myers R C 2003 {\it JHEP } {\bf 0309} 025 

\bibitem{hawking-page} Hawking S W and Page D N 1983 {\it Commun. Math. Phys.} {\bf 87} 577

\bibitem{blackring} Emparan R and Reall H S 2002 {\it Phys. Rev. Lett.} \textbf{88}  101101

\item[] Elvang H, Emparan R, Mateos D and Reall H S 2004
{\em Phys. Rev. Lett.} {\bf 93} 211302

\bibitem{idealgas} Nulton J D and Salamon P 1985 {\it Phys. Rev. A} {\bf 31} 2520

\bibitem{classi} \AA man J E 2002 {\it Manual for CLASSI: Classification Programs for Geometries
in General Relativity}, Department of Physics, Stockholm University, Technical Report, Provisional Edition,
Distributed with the sources for SHEEP and CLASSI 

\bibitem{grtensor} GRTensor II Version 1.79 (R4) 2001 {\it http://grtensor.phy.queensu.ca/}  

\end{thebibliography}
\end{document}